\documentclass[a4paper,10pt]{amsart}

\usepackage{amssymb,amsmath,amsthm,ascmac,array,bm}

\usepackage{graphicx}
\usepackage{xcolor}

\usepackage{algorithmicx,algorithm}
\usepackage[noend]{algpseudocode}
\usepackage{enumitem}
\usepackage{booktabs}  
\usepackage{threeparttable,multirow,array} 
\usepackage[most]{tcolorbox}
\usepackage{slashbox} 
\usepackage{soul}
\usepackage{bbm}

\usepackage{mathtools}

\usepackage{fancyvrb}

\mathtoolsset{showonlyrefs} 

\usepackage{tikz}
\usetikzlibrary{positioning,shapes,shadows,arrows}

\numberwithin{equation}{section}
\allowdisplaybreaks

\newcommand{\mcl}{\mathcal{L}}

\newcommand{\mbbh}{\mathbb{H}}

\newcommand{\mbbr}{\mathbb{R}}

\newcommand{\al}{\alpha}  \newcommand{\ep}{\epsilon} 
\newcommand{\vp}{\varphi} \newcommand{\del}{\delta} \newcommand{\sig}{\sigma}
\newcommand{\D}{\Delta}  \newcommand{\gam}{\gamma}
 \newcommand{\Gam}{\Gamma}

\newcommand{\cil}{\xrightarrow{\mcl}} 

\newcommand{\argmax}{\mathop{\rm argmax}}

\def\nn{\nonumber}

\def\tcb#1{\textcolor{black}{#1}}

\def\diag{{\rm diag}}

\def\lp{L\'evy process}

\def\cadlag{c\`adl\`ag}

\newcommand{\nes}{\hat{\nu}_{n}}

\newcommand{\mes}{\hat{\mu}_{n}}
\newcommand{\ses}{\hat{\sig}_{n}}
\newcommand{\hep}{\hat{\epsilon}}

\newcommand{\sumi}{\sum_{i=1}^{[T_n]}}

\newcommand{\yuima}{YUIMA }

\DefineVerbatimEnvironment{Sinput}{Verbatim}{fontshape=n}
\DefineVerbatimEnvironment{Soutput}{Verbatim}{}
\DefineVerbatimEnvironment{Scode}{Verbatim}{fontshape=n}
\DefineVerbatimEnvironment{Code}{Verbatim}{}
\DefineVerbatimEnvironment{CodeInput}{Verbatim}{fontshape=n}
\DefineVerbatimEnvironment{CodeOutput}{Verbatim}{}
\newenvironment{CodeChunk}{}{}
\setkeys{Gin}{width=0.8\textwidth}

\let\proglang=\textsf
\let\code=\texttt

\title[Student $t$-L\'{e}vy regression model in \yuima]{Student $t$-L\'{e}vy regression model in \yuima}

\author[H. Masuda]{Hiroki Masuda}
\address{Graduate School of Mathematical Sciences,  University of Tokyo, Japan
}
\email{hmasuda@ms.u-tokyo.ac.jp}

\author[L. Mercuri]{Lorenzo Mercuri}
\address{Department of Economics, Management and Quantitative Methods, University of Milan, Italy\\
INDAM-GNAMPA: Gruppo Nazionale per l'Analisi Matematica, la Probabilit\'a  e le loro Applicazioni\\
}
\email{lorenzo.mercuri@unimi.it}

\author[Y. Uehara]{Yuma Uehara}
\address{Department of Mathematics, Faculty of Engineering Science, Kansai University, Japan}
\email{y-uehara@kansai-u.ac.jp}

\date{\today}
\begin{document}
\maketitle

\begin{abstract}
The aim of this paper is to discuss an estimation and a simulation method in the \textsf{R} package \yuima for a linear regression model driven by a Student-$t$ L\'evy process with constant scale and arbitrary degrees of freedom. This process finds applications in several fields, for example finance, physic, biology, etc.
The model presents two main issues. The first is related to the simulation of a sample path at high-frequency level. Indeed, only the $t$-L\'evy increments defined on an unitary time interval are Student-$t$ distributed. In \yuima, we solve this problem by means of the inverse Fourier transform for simulating the increments of a Student-$t$ L\'{e}vy defined on a interval with any length. A second problem is due to the fact that joint estimation of trend, scale, and degrees of freedom does not seem to have been investigated as yet. In \yuima, we develop a two-step estimation procedure that efficiently deals with this issue. Numerical examples are given in order to explain methods and classes used in the \yuima package.
\end{abstract}


%
%
%
%

\section{Introduction \label{Intro}}

\yuima package in R provides several simulation and estimation methods for stochastic processes \tcb{\cite{JSSv057i04,iacus2018simulation}}.
This paper introduces new classes and methods in \yuima for simulating and estimating a $t$-L\'evy regression model based on high-frequency observations. This model can be seen as a generalization of a $t$-L\'evy process (cf. \cite{HeyLeo05,Cuf07}) by adding covariates, which can be either deterministic or stochastic processes.
Adding covariates to a continuous-time stochastic process is a widely used approach for constructing new processes in various fields. For example, in finance, periodic deterministic covariates can be employed to capture the seasonality observed in commodity markets \cite{sorensen2002modeling}. Alternatively, in insurance and medicine, covariates are often incorporated into mortality rate dynamics to account for age and cohort effects (see \cite{HABERMAN2009255,castro2021trends} and references therein).


In such fields, data often exhibit heavy-tailed behavior in the margins and thus, the inclusion of a $t$-L\'evy process as a driving noise would be useful.
However, despite the simple mathematical definition of a $t$-L\'evy regression model, it poses significant challenges both in deriving its mathematical properties and in implementing numerical methods. A key difficulty arises from the fact that the $t$-L\'evy driving noise is not closed under convolution. Consequently, the distribution of its increments follows a $t$-distribution only over a unit-time interval.

Motivated by this fact, in \yuima, we propose three simulation methods for a $t$-L\'evy regression model.
The key element is the construction of a random number generator for the noise increments, which is essential for several discretized simulation methods and involves the numerical inversion of the characteristic function.
More specifically, our method is based on the approximation of the cumulative distribution function, and our method can relieve numerical instability compared with directly using the density function especially for simulating small time increments.
Additionally, \yuima provides a two-stage estimation procedure and the corresponding estimator has consistency and asymptotic normality as shown in \cite{masuda2023quasilikelihood}.

The rest of this paper is organized as follows. We briefly summarize the $t$-L\'evy regression model in Section \ref{Math:Model}. We introduce the new \yuima classes and methods in Section \ref{Sect:N1} and we show some examples with simulated and real \tcb{data} that highlight their usage in Section \ref{sec:NE}. Finally, Section \ref{concl} concludes the paper.

\section{t-L\'evy regression model \label{Math:Model}}

In this section, we review the main characteristics of the $t$-L\'evy regression model and the $t$-L\'evy process used as its driving noise. We highlight the main issues that arise in the simulation and estimation procedures for these models. For a complete discussion on all mathematical aspects and the properties of these procedures in both cases, we refer to \cite{masuda2023quasilikelihood}.

\tcb{The proposed} $t$-L\'evy regression model is a continuous-time stochastic process of the form: 
\begin{equation}
Y_t = X_t \cdot \mu + \sig J_t, \qquad t\in[0,T_n],
\label{hm:model}
\end{equation}
where the $q$-dimensional vector process $X=(X_t)$ with {\cadlag} paths contains the covariates; the dot denotes the inner product in $\mbbr^q$ and $J=(J_t)$ is a $t$-{\lp} such that its unit-time distribution $\mcl(J_1)$ is given by:
\begin{equation}
\mcl(J_1) = t_\nu:=t_\nu(0,1),
\label{hm:t_nu(0,1)}
\end{equation}
where $t_\nu(\mu_1,\sig)$ denotes the scaled Student-$t$ distribution \tcb{with density:
\begin{equation*}
f(x;\mu,\sig,\nu) := \frac{\Gamma(\frac{\nu+1}{2})}{\sig\sqrt{\pi}\Gamma(\frac{\nu}{2})}
\left\{1+\left(\frac{x-\mu}{\sig}\right)^2\right\}^{-(\nu+1)/2}.
\end{equation*}}
The parameters $\nu>0$, $\mu_1 \in \mbbr$ and $\sig>0$ represent the degree of freedom, position, and scale parameters, respectively (see \cite{HeyLeo05,Cuf07} for more details on a $t$-{\lp}).
For this model, we consider the situation where we estimate these three unknown parameters based a discrete-time sample $\{(X_{t_j},Y_{t_j})\}_{j=0}^{[nT_n]}$ with $t_j=t_j^n:=\frac{j}{n}$ and $T_n\to \infty$ as $n\to\infty$.

To develop the \yuima simulation and estimation algorithms for the model in \eqref{hm:model}, we need to address two problems: first, the simulation of the sample path of $J=(J_t)$ on a small time grid with $\Delta t \neq1$ which is essential for \tcb{handling the increments: 
$$\D_j Y = \D_j X \cdot \mu +\sig \D_j J, \quad j=0,1,\dots, [nT_n]$$}
where $\D_j Z$ denotes $Z_{t_j}-Z_{t_{j-1}}$ for any stochastic process $Z=(Z_t)$.
Second, the identification of an efficient procedure for estimating the model parameters $\left(\mu, \sig \right)$ in \eqref{hm:model} and the degree of freedom $\nu$ in \eqref{hm:t_nu(0,1)}.
We will introduce both the simulation and estimation \tcb{methods} below.


Due to the stationary behavior of the L\'evy increments, i.e. $\mcl(\Delta_i J) = \mcl(J_h)$ and $h:=t_i-t_{i-1}=\frac{1}{n}$ for $i = 1,\ldots, [nT_n]$, $\mcl(J_h)$ admits the Lebesgue density:
\begin{align}
x &\mapsto 
\frac{1}{\pi}\int_0^\infty \cos(ux)\{\vp_{J_1,\nu}(u)\}^h du
\nn\\
&= \left(\frac{2^{1-\nu/2}}{\Gam(\nu/2)} \right)^h
\frac{1}{\pi}\int_0^\infty \cos(ux) u^{\nu h/2} \left(K_{\nu/2}(u)\right)^h du,
\label{hm:ff_def_pre}
\end{align}
where  $\vp_{J_1,\nu}(u)$, the characteristic function of $\mcl(J_1)=t_\nu$ is given by
\begin{equation}
\vp_{J_1,\nu}(u) := \frac{2^{1-\nu/2}}{\Gam(\nu/2)} |u|^{\nu/2} K_{\nu/2}(|u|), \qquad u\in\mbbr
\label{hm:t_nu(0,1)-CF}
\end{equation}
and $K_\nu(t)$ denotes the modified Bessel function of the second kind ($\nu\in\mbbr$, $t>0$):
\begin{equation}
K_{\nu}(t)=\frac{1}{2}\int_0^{\infty}s^{\nu-1}\exp\left\{-\frac{t}{2}\left(s+\frac{1}{s}\right)\right\}ds.
\nonumber
\end{equation}
We here remark that although the explicit expression of \eqref{hm:ff_def_pre} cannot be obtained for general $h$, the tail index of $\mathcal{L}(J_h)$ is the same as that of $\mathcal{L}(J_1)$ (cf. \cite{BerVig08} Theorem 2), and hence, modeling tail index via student $t$-L\'evy process makes sense. 
A classical method for simulating $t$-L\'evy increments based on this Fourier inversion representation is through the rejection method, as discussed in \cite{Hubalek2005OnSF} and \cite{Dev81}. However, this method can become time-consuming and unstable when dealing with very small values of $h$, primarily due to the oscillatory behavior encountered during the numerical evaluation of density.
Such a problem often arises since to express the high-frequently observed situation from a continuous process, we should take a finer mesh of simulating the underlying process than that of simulating the discrete observations.
For this issue, in \yuima, the Random Number Generator for the increments is developed using the inverse quantile method. The first step involves integrating the density in \eqref{hm:ff_def_pre} to obtain the cumulative distribution function (CDF). This approach leads to a more stable behavior in the tails of the CDF compared to the density tails, owing to a mitigation effect observed during the numerical evaluation of the following double integral:
\begin{equation}
F(y) =\int_{-\infty}^{y} \frac{1}{\pi}\int_0^\infty \cos(ux)\{\vp_{J_1,\nu}(u)\}^h du dx.
\end{equation}    
Further details on our approach are provided in the \tcb{next} section.
Another way to simulate $t$-L\'evy process is using series representation. Numerically, for each time $t$, we need ad-hoc truncation of the infinite sum. The Gaussian approximation of a small-jump part is also valid  \cite{AsmRos01}, but the associated error may not be easy to control in a practical manner.
For more details, see \cite{Mass18}.

Following \cite{masuda2023quasilikelihood}, \yuima provides a two-stage estimation algorithm for $\left(\mu, \sig, \nu\right)\in \Theta_\mu\times\Theta_\sig \times \Theta_\nu$. 
Denote by $(\mu_0,\sig_0,\nu_0)$ the parameter true values of the model in \eqref{hm:model}; 
the two-step algorithm can be summarized as follows:
\begin{enumerate}
\item An estimator $\hat{a}:=(\mes,\ses)$ of $a_0=(\mu_0,\sig_0)$ is obtained by maximizing the \textit{Cauchy quasi-likelihood}:
\begin{equation}
\hat{a}_n:=(\mes,\ses)\in\argmax_{a\in \overline{\Theta_\mu\times\Theta_\sig}}\mbbh_{1,n}(a).
\label{hm:def_CQMLE}
\end{equation}
The Cauchy quasi-(log-)likelihood $\mbbh_{1,n}(a)$ conditional on $X$ has the following form:
\begin{align}
\mbbh_{1,n}(a) 
&:= \sum_{j=1}^{N_n} \log\left\{\frac{1}{h\sig}\phi_1\left( \frac{\D_j Y - \mu\cdot \D_j X}{h\sig} \right)\right\}
\nn\\
&= C_n - \sum_{j=1}^{N_n} \left\{ \log\sig + \log\left(1+\ep_j(a)^2\right)\right\},
\nonumber
\end{align}
where $\phi_1$ denotes the density function of the standard Cauchy distribution and the term $C_n$ does not depend \tcb{on $a=\left(\mu,\sigma\right)$}. $N_n:=[n B_n]$ is the number of observations in the part $[0,B_n]$ of the entire period $[0, T_n]$ where $(B_n)$ is a positive sequence satisfying $B_n\le T_n$ and $\qquad n^{\ep''} \lesssim B_n \lesssim n^{1-\ep'}$ for some $\ep',\ep''\in(0,1)$. 
\item Then, we construct an estimator $\nes$ of $\nu_0$ using the \textit{Student-$t$ quasi-likelihood} on the ``unit-time'' residual sequence $\hep_i$:
\begin{equation}
\hep_i := \ses^{-1}\left(Y_i - Y_{i-1} - \mes\cdot(X_i - X_{i-1})\right),
\nonumber
\end{equation}
for $i=1,\dots,[T_n]$, which is expected to be approximately i.i.d. $t_\nu$-distributed. Therefore $\nes$ solves:
\begin{equation}
\nes\in\argmax_{\nu\in\overline{\Theta}_\nu}\mbbh_{2,n}(\nu),
\nonumber
\end{equation}
\tcb{where}
\begin{equation}
\mbbh_{2,n}(\nu) := 
\sumi\left( -\frac12 \log\pi 
+ \log\Gam\left(\frac{\nu+1}{2}\right) - \log\Gam\left(\frac{\nu}{2}\right)
- \frac{\nu+1}{2}\log\left( 1+ \hep_i^2\right)
\right).
\nonumber
\end{equation}
\end{enumerate}

We note that the first estimation scheme is based on the locally Cauchy property of $J$: $h^{-1}J_h\cil t_1$ (standard Cauchy) \tcb{as $h \to 0$}.
Let \tcb{$\hat{u}_{a,n}:=\sqrt{N_n}(\hat{a}_n-a_0)$ and $\hat{u}_{\nu,n}:=\sqrt{T_n}(\hat{\nu}_n-\nu_0)$.
Under some regularity conditions on the covariate process} $X=\left(X_t\right)$ (see Assumption 2.1 in \cite{masuda2023quasilikelihood}for all requirements on $X$), the estimators have the following joint asymptotic normality:
$$(\hat{\Gamma}_{a,n}^{1/2}\hat{u}_{a,n}, \hat{\Gamma}_{\nu,n}^{1/2}\hat{u}_{\nu,n})\cil N_{q+2}(0,I_{q+2}),$$
where $\psi_1$ denotes the trigamma function and
\begin{align*}
&\hat{\Gamma}_{a,n}=\diag \left(\frac{1}{2\hat{\sig}_n^2N_n}\sum_{j=1}^{N_n} \left(\frac{1}{h}\D_j X\right)^{\otimes2}, \frac{1}{2\hat{\sig}_n^2}\right),\\
& \hat{\Gamma}_{\nu,n}=\frac{1}{4}\Bigg(\psi_1\Bigg(\frac{\hat{\nu}_n}{2}\Bigg)-\psi_1\Bigg(\frac{\hat{\nu}_n+1}{2}\Bigg)\Bigg).
\end{align*}
Since $\hat{\Gamma}_{a,n}^{1/2}$ and $\hat{\Gamma}_{\nu,n}$ can be constructed only by the observations, we can easily obtain the confidence intervals of each parameter.

\section{Classes and Methods for t-L\'evy Regression Models \label{Sect:N1}}


This section provides an overview of the new classes and methods introduced in \yuima for the mathematical definition, trajectory simulation, and estimation of a Student L\'evy Regression model. To handle this model, the first step involves constructing an object of the \code{yuima.LevyRM-class}. As an extension of the \code{yuima-class} (refer to \cite{JSSv057i04} for more details), \code{yuima.LevyRM-class} inherits slots such as \code{@data}, \code{@model}, \code{@sampling}, and \code{@functional} from its parent class. The remaining slots store specific information related to the Student-$t$-L\'evy regression model. 

Notably, the slot \verb|@unit_Levy| contains an object of the \code{yuima.th} class, which represents the mathematical description of the Student-$t$ L\'evy process $J_t$ (see the subsequent section for detailed explanations). The labels of the regressors are saved in the slot \code{@regressors}, while slots \code{@LevyRM} and \code{@paramRM} respectively cache the names of the output process $Y_t$ and a \code{string} vector reporting the regressors' coefficients, the scale parameter, and the degree of freedom. The \code{yuima.th-class} is obtained by the new \verb|setLaw_th| constructor. This function requires the arguments used for the numerical inversion of the characteristic function, and its usage is discussed in the next section. 

Once the \code{yuima.th-object} is created, we define the system of stochastic differential equations (SDEs) that describes the behavior of the regressors, with their mathematical definitions stored in an object of the \code{yuima.model-class}. Both \code{yuima.th} and \code{yuima.model} objects are used as inputs for the \code{setLRM} constructor, which returns an object of the \code{yuima.LevyRM-class}. The following chunk code reports the input for this new function.
\begin{CodeChunk}
\begin{CodeInput}
setLRM(unit_Levy, yuima_regressors, LevyRM = "Y", coeff = c("mu", "sigma0"), data = NULL,
  sampling = NULL, characteristic = NULL, functional = NULL)
\end{CodeInput}
\end{CodeChunk}
As is customary for any class extending the \code{yuima-class}, the \code{simulate} method enables the generation of sample paths for the Student L\'evy Regression model. To simulate trajectories, an object of the \code{yuima.sampling-class} is constructed to represent an equally spaced grid-time used in trajectory simulation. The regressors' paths are obtained using the Euler scheme, while the increments of the Student-$t$ L\'evy process are simulated using the random number generator available in the slot \code{@rng} of the \code{yuima.th-object}.

The last method available in \yuima is \verb|estimation_LRM|. This method allows the users to estimate the model using either real or simulated data. The estimation follows a two-step procedure, introduced \tcb{in \cite{masuda2023quasilikelihood}}.
%
%
%
\begin{CodeChunk}
\begin{CodeInput}
estimation_LRM(start, model, data, upper, lower)
\end{CodeInput}
\end{CodeChunk}
For this function, the minimal inputs are \code{start}, \code{model}, \code{data}, \code{upper} and \code{lower}. The arguments \code{start} are the initial points for the optimization routine while \code{upper} and \code{lower} corresponds to the box constraints. The \code{yuima.LevyRM-object} is passed to the function through the input \code{model} while the input \code{data} is used to pass the dataset to the internal optimization routine.  
Figure \ref{tLRm} describes these new classes and methods, along with their respective usage.
\pgfdeclarelayer{background}
\pgfdeclarelayer{foreground}
\pgfsetlayers{background,main,foreground}
\tikzstyle{sensor}=[draw, fill=gray!10, text width=7.5em, 
    text centered, minimum height=3em,drop shadow]
\tikzstyle{sensor1}=[draw, diamond, aspect=2, fill=gray!10, text width=7.5em, 
    text centered, minimum height=3em,drop shadow]
\tikzstyle{sensor2}=[draw, ellipse, aspect=2, fill=gray!10, text width=7.5em, 
    text centered, minimum height=3em,drop shadow]
\def\blockdist{2.3}
\def\edgedist{2.5}
\begin{figure}[!ht]
\centering
\begin{tikzpicture}[scale=0.60, transform shape]
    \node (pos1) [sensor]  {\code{yuima.LevyRM}};
		\path (pos1)+(-6.1,0) node (cons) [sensor1]{\code{setLRM}};
		\draw[<-] (pos1) edge (cons);
		\path (cons)+(-6.1,0) node (cons1) [sensor]{\code{yuima.th}};
		\draw[<-] (cons) edge (cons1);
		\path (cons1)+(0,-1.5) node (cons2) [sensor]{\code{yuima.model}};
    \draw[->](cons2) -| (cons);
		\path (cons1)+(-6.1,0) node (cons3) [sensor1]{\verb|setLaw_th|};
		\draw[->] (cons3) edge (cons1);
		\path (cons3)+(0,3) node (cons4) [sensor]{Inputs for the Fourier Inv.};
		\draw[->] (cons4) edge (cons3);
		\path (cons)+(0,-3) node (cons5) [sensor1]{\code{simulate}};
		\draw[->] (pos1) |- (cons5);
		\path (cons2)+(0,-1.5) node (cons6) [sensor]{\code{yuima.sampling}};
		\draw[->] (cons6) edge (cons5);
		\path (cons5)+(0,-3) node (cons7) [sensor2]{\code{Sample Path}};
		\draw[->] (cons5) edge (cons7);
		\path (cons)+(0,3) node (cons8) [sensor1]{\verb|estimation_LRM|};
		\draw[->] (pos1) |- (cons8);
		\path (cons1)+(0,3) node (cons9) [sensor]{\code{yuima.data}};
		\draw[->] (cons9) edge (cons8);
		\path (cons8)+(0,3) node (cons10) [sensor2]{\code{Estimated Param.}};
		\draw[->] (cons8) edge (cons10);
\end{tikzpicture}
\caption{Scheme of classes and methods in \yuima for the t-L\'evy Regression model. \label{tLRm}}
\end{figure}
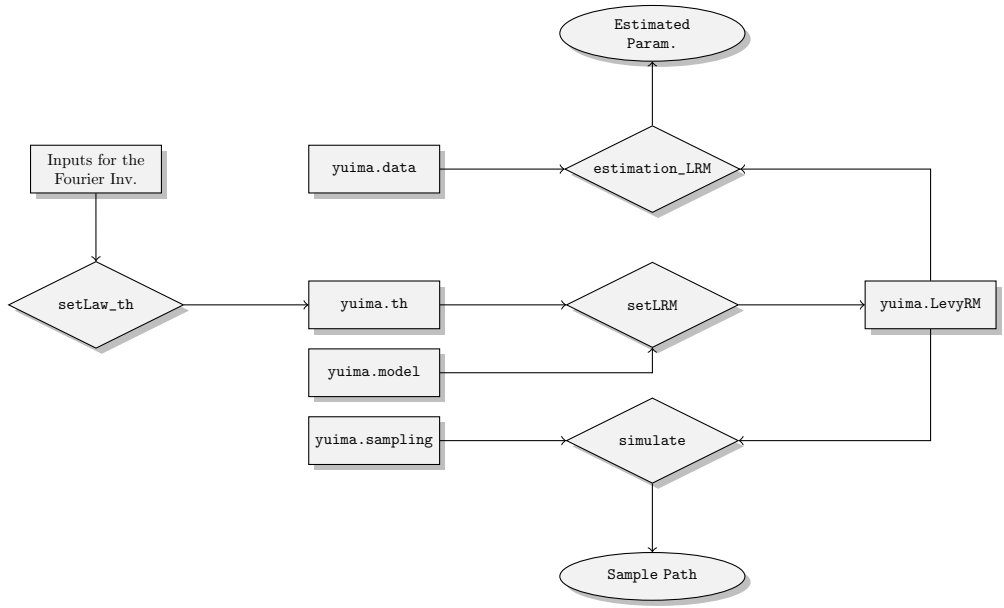

\subsection{yuima.th: A new class for mathematical description of a Student-t L\'evy process \label{yuimath}}

In this section, the steps for the construction of an \code{yuima.th-object} are presented. As remarked in Section \ref{Sect:N1}, this object contains all information on the Student-$t$ L\'evy process. Moreover, detailed information is provided regarding the numerical algorithms utilized for evaluating the density function \eqref{hm:ff_def_pre}. The construction of this object is accomplished using the \verb|setLaw_th| constructor, and the subsequent code snippet displays its corresponding arguments.
\begin{CodeChunk}
\begin{CodeInput}
setLaw_th(h = 1, method = "LAG", up = 7, low = -7, N = 180, N_grid = 1000, 
  regular_par = NULL)
\end{CodeInput}
\end{CodeChunk}
The input \code{h} is the length of the step-size of each time interval for the Student-$t$ L\'evy increment $\D J_h=J_h-J_0$. Its default value, \code{h=1}, indicates that the \code{yuima.th-object} describes completely the process $J_t$ at time 1. The argument \code{method} refers to the type of quadrature used for the computation of the integral in \eqref{hm:ff_def_pre} while the remaining arguments govern the precision of the integration routine. 

The \code{yuima.th-class} inherits the slots \code{@rng}, \code{@density}, \code{@cdf} and \code{@quantile} from its parent class \code{yuima-law} \cite{Masuda2022}. These slots store respectively the random number generator, the density function, the cumulative distribution function and the quantile function of $\Delta J_h$. \newline  As mentioned in Section \ref{Intro}, the density function with $h\neq1$ does not have a closed-form formula and, therefore, the inversion of the characteristic function is necessary. \yuima provides three methods for this purpose: the Laguerre quadrature, the COS method and the Fast Fourier Transform. 
The Gauss-Laguerre quadrature is a numerical integration method employed for evaluation integrals in the following form:
\[
\mathcal{I}=\int_0^{+\infty} f\left(x\right) e^{-x}\mbox{d}x.
\]
This procedure has been recently used for the computation of the density of the variance gamma and the transition density of a CARMA model driven by a time-changed Brownian motion \cite{loregian2012approximation,mercuri2021finite}, this motivates its application in this paper. 

Let $f\left(x\right)$ be a continuous function defined on $\left[0, +\infty\right)$ such that:
\begin{equation*}
\mathcal{I}=\int_0^{+\infty} f\left(x\right) e^{-x}\mbox{d}x<+\infty;
\end{equation*}
the integral $\mathcal{I}$ can be approximated as follows:
\begin{equation}
\mathcal{I}\approx \sum_{j=1}^{N}\omega\left(k_j\right)f\left(k_{j}\right),
\label{eq:Laguerre}
\end{equation}
where $k_j$ is the $j$th-root of the $N$-order Laguerre polynomial\footnote{The Laguerre polynomial can be defined recursively as follows: $L_0\left(x\right), L_1(x)=1-x, \ldots, L_{N}\left(x\right)=\frac{\left(2N-x\right)L_{N-1}\left(x\right)-\left(N-1\right)L_{N-2}\left(x\right)}{N}$}  $L_N\left(x\right)$ and the weights $\omega\left(k_j\right), j=1,\ldots, N$ are defined as:
\begin{equation}
\omega\left(k_{j}\right) = \frac{k_{j}}{\left(N+1\right)^2 L_{N+1}^2\left(k_j\right)}.
\end{equation}
To apply the approximation in \eqref{eq:Laguerre}, we rewrite the inversion formula \tcb{in \eqref{hm:ff_def_pre}} as follows:
\begin{eqnarray}
f\left(x\right)&=&\frac{1}{\pi}\left(\frac{2^{1-\frac{\nu}{2}}}{\Gamma\left(\nu/2\right)}\right)\int_{0}^{+\infty}\cos\left(ux\right)u^{\nu h/2}\left(K_{\nu/2}\left(u\right)\right)^h\mbox{d}u\nonumber\\
&=&\frac{1}{\pi}\left(\frac{2^{1-\frac{\nu}{2}}}{\Gamma\left(\nu/2\right)}\right)\int_{0}^{+\infty}\cos\left(ux\right)u^{\nu h/2}\left(K_{\nu/2}\left(u\right)\right)^he^{u}e^{-u}\mbox{d}u.
\label{eq:InvForLM}
\end{eqnarray} 
Applying the result in \eqref{eq:Laguerre}, the density function \tcb{of $J_h$} can be approximated with the formula reported below:
\begin{equation*}
\hat{f}_{N}\left(x_j\right) = \frac{1}{\pi}\left(\frac{2^{1-\frac{\nu}{2}}}{\Gamma\left(\nu/2\right)}\right) \sum_{j=1}^{N}\cos\left(k_j x\right)k_j^{\nu h/2}\left(K_{\nu/2}\left(k_j\right)\right)^he^{k_j}\omega\left(k_j\right).
\label{eq:FinalLaguerre}
\end{equation*}
Notably, the approximation formula's precision in equation \eqref{eq:FinalLaguerre} can be enhanced through the argument \code{N} in the \verb|setLaw_th| constructor.  The roots $k_j$ and the weights $\omega\left(k_j\right)$ are internally computed using the \code{gauss.quad} function from the \proglang{R} package \code{statmod}, with a maximum allowed order of 180 for the Laguerre polynomial.

The COS method is based on the Fourier Cosine expansion employed for an even function with the compact domain $\left[-\pi, \pi \right]$. This method has been widely applied in the finance literature for the computation of the exercise probability of an option and its no-arbitrage price, we refer to \cite{fang2009novel,fang2011fourier} and references therein for details. Let $g\left(\theta\right):\left[-\pi, \pi \right] \rightarrow \mathbb{R}$ be an even function, its Fourier Cosine expansion reads:
\begin{equation}
g\left(\theta\right)=\frac12 a_0 +\sum_{k=1}^{\infty} a_k \cos\left(k\theta\right)
\label{eq:ExpCosF}
\end{equation} 
with
\begin{equation*}
a_k = \frac{2}{\pi}\int_0^{\pi}g\left(\theta\right)\cos\left(k\theta\right)\mbox{d}\theta.
\end{equation*}
Denoting with $f\left(x\right)$ the density function of the increment $\Delta J_h$, the new function $\bar{g}\left(\theta\right)$ is defined as:
\begin{equation}
\bar{g}\left(\theta\right):=f\left(\frac{L}{\pi}\theta\right)\frac{L}{\pi} \mathbbm{1}_{\left\{-\pi \leq \theta \leq \pi\right\} },
\label{eq:newg}
\end{equation}
The function $\bar{g}\left(\theta\right)$ is still an even function for any $L>0$ and the Fourier Cosine expansion in \eqref{eq:ExpCosF} can be applied. The coefficient $a_k$ is determined as follows:
\begin{eqnarray*}
a_k &=& \frac{2}{\pi}\int_0^{\pi}\bar{g}\left(\theta\right) \cos\left(k\theta\right)\mbox{d}\theta\nonumber\\
&=&\frac{2}{\pi}\int_0^{\pi} f\left(\frac{L}{\pi}\theta\right)\cos\left(k\theta\right)\frac{L}{\pi}\mbox{d}\theta.\nonumber
\end{eqnarray*}
Setting $\theta = \frac{\pi}{L}x$, we have:
\begin{equation}
a_k = \frac{2}{\pi}\int_0^{L}f\left(x\right) \cos\left(k\frac{\pi}{L}x\right)\mbox{d}x.
\end{equation}
The coefficient $a_k$ can be rewritten using the characteristic function of $\Delta J_h$ at $k\frac{\pi}{L}$, i.e.:
\begin{equation}
a_k = \frac{2}{\pi}\left[\varphi\left(k\frac{\pi}{L}\right)-\int_{L}^{+\infty}f\left(x\right) \cos\left(k\frac{\pi}{L}x\right)\mbox{d}x\right].
\end{equation}
For a sufficiently large $L$, the coefficient $a_k$ can be approximated as follows:
\begin{equation}
a_k \approx \frac{2}{\pi} \varphi\left(k\frac{\pi}{L}\right).
\end{equation}
The following series expansion is achieved:
\begin{equation}
f\left(x\right)=\frac12 \bar{a}_0 + \sum_{k=1}^{+\infty} \bar{a}_k\cos\left(k\frac{\pi}{L}x\right)
\label{CosMethodInYuima}
\end{equation}
with
\[
\bar{a}_k = a_k \frac{\pi}{L} \approx \frac{2}{L}\varphi\left(k\frac{\pi}{L}\right).
\]
Finally, \eqref{CosMethodInYuima} can be approximated by truncating the series as follows:
\begin{equation}
\hat{f}_{N}\left(x\right)=\frac12 \hat{a}_{0,L} + \sum_{k=1}^{N} \hat{a}_{k,L}\cos\left(k\frac{\pi}{L}x\right)
\label{ApproxCosMethodInYuima}
\end{equation}
with 
\[
\hat{a}_{k,L}=\frac{2}{L}\varphi\left(k\frac{\pi}{L}\right).
\]
The precision of $\hat{f}_{N,L}\left(x\right)$ in \eqref{ApproxCosMethodInYuima} depends on $N$ and $L$. Users can select these two quantities using \code{N} and the couple (\code{up}, \code{low}) respectively. $L$ is computed internally in the \verb|setLaw_th| constructor as $\code{L = max(|low|,up)}$. 

The last method available in \yuima is the Fast Fourier Transform (\texttt{FFT}) \cite{singleton1969algorithm,cooley1965algorithm} for the inversion of the characteristic function. \yuima uses the \code{FFT} method developed internally in the function \verb|FromCF2yuima_law| for the inversion of any characteristic function defined by users. In the latter case, the density function of $\Delta J_h$ is based on the following general inversion formula:
\begin{equation}
f\left(x\right) = \frac{1}{2\pi}\int_{-\infty}^{+\infty} e^{-iux}\varphi\left(u\right) \mbox{d}u.
\label{eq:INVCharLM}
\end{equation} 
As first step, we apply to the integral in \eqref{eq:INVCharLM} the change variable $u = 2\pi \omega$ and we have:
\begin{equation}
f\left(x\right) = \int_{-\infty}^{+\infty} e^{-i 2\pi\omega x}\varphi\left(2\pi\omega\right) \mbox{d}\omega.
\end{equation}
We consider discrete support for $x$ and $\omega$. The $x$ grid has the following structure:
\begin{equation}
x_0 =a, \ldots, x_j=a+ j \Delta x, \ldots, x_N=b
\label{grid_x}
\end{equation}
with $\Delta x =\frac{b-a}{N}$ where $N+1$ is the number of the points in the grid in \eqref{grid_x}. 
Similarly, we define a $\omega$ grid:
\begin{equation}
\omega_0= -\frac{N}{2} \Delta \omega, \ldots,  \omega_n = -\frac{N}{2} + n \Delta \omega  \ldots, \omega_N = \frac{N}{2}\Delta \omega
\label{omega_grid}
\end{equation}
where $\Delta \omega = \frac{1}{N \Delta x} = \frac{1}{b-a}$. Both grids have the same dimension $N+1$, $\Delta x$ shrinks as $N \rightarrow +\infty$ while $\Delta \omega$ reduces for large values of $b-a$.
For  any $x_j$ in the grid we can approximate the integral in \eqref{eq:INVCharLM} using the left Riemann summation, therefore, we have the approximation $\hat{f}_{N}\left(x_j\right)$:
\begin{align}
\hat{f}_{N}\left(x_j\right) &= \Delta \omega \sum_{n=1}^{N}e^{2 i\pi \left(-\frac{N}{2}\Delta \omega +\left(n-1\right) \Delta u\right)x_j}\varphi\left(-\pi N\Delta u + 2\pi \left(n-1\right)\right)\nn\\
&= \Delta \omega e^{i\pi N \Delta \omega} \sum_{n=1}^{N}e^{- 2 i \pi \left(n-1\right)\Delta u \left(a+\left(j-1\right) \Delta x\right)}\varphi\left(-\pi N\Delta \omega + 2\pi \left(n-1\right)\right)\nn\\
&= \Delta \omega e^{i\pi N \Delta \omega}\sum_{n=1}^{N}e^{\frac{- 2 i \pi\left(n-1\right)\left(j-1\right)}{N}}\varphi\left(-\pi N\Delta \omega + 2\pi \left(n-1\right)\right)e^{\frac{-2i\pi \left(n-1\right)a}{N}}.
\label{FFT_Dens}
\end{align}
The last equality in \eqref{FFT_Dens} is due to the identity $\Delta \omega \Delta x = \frac{1}{N}$. To evaluate the summation in \eqref{FFT_Dens}, we use the $\texttt{FFT}$ algorithm and
\begin{equation}
\hat{f}_{N}\left(x_j\right) = \Delta \omega e^{i\pi N \Delta \omega}\texttt{FFT}\left[\varphi\left(-\pi N\Delta \omega + 2\pi \left(n-1\right)\right)e^{ \frac{-2 i \pi \left(n-1\right)a}{N}}\right].
\end{equation}
In this case, we have two sources of approximation errors that we can control using the arguments \code{up}, \code{low} and \code{N} in the \verb|setLaw_th| constructor. \code{N} denotes the number of intervals in the grid used for the Left-Riemann summation while \code{up}, \code{low} are used to compute the step size of this grid. 

Once the density function has been obtained using one of the three methods described above, it is possible to approximate the cumulative distribution function using the Left-Riemann summation computed on the grid in \eqref{grid_x}, therefore the cumulative distribution function $F\left(\cdot\right)$ for each $x_j$ on the grid is determined as follows:
\begin{equation}
\hat{F}\left(x_j\right) = \sum_{x_k<x_j}\hat{f}_{N}\left(x_k\right)\Delta x. 
\end{equation}
In this way, we can construct a table that we can use internally in the \code{cdf} and \code{quantile} functions. Moreover, to evaluate the cumulative distribution function at any $x\in \left( x_{j-1}, x_j \right)$, we  interpolate linearly its value using the couples $\left(x_{j-1}, \hat{F}\left(x_{j-1}\right)\right)$ and $\left(x_{j}, \hat{F}\left(x_{j}\right)\right)$. The random numbers can be obtained using the inversion sampling method.

\begin{figure}[!htbp]
	\centering
		\includegraphics[width=0.75\textwidth]{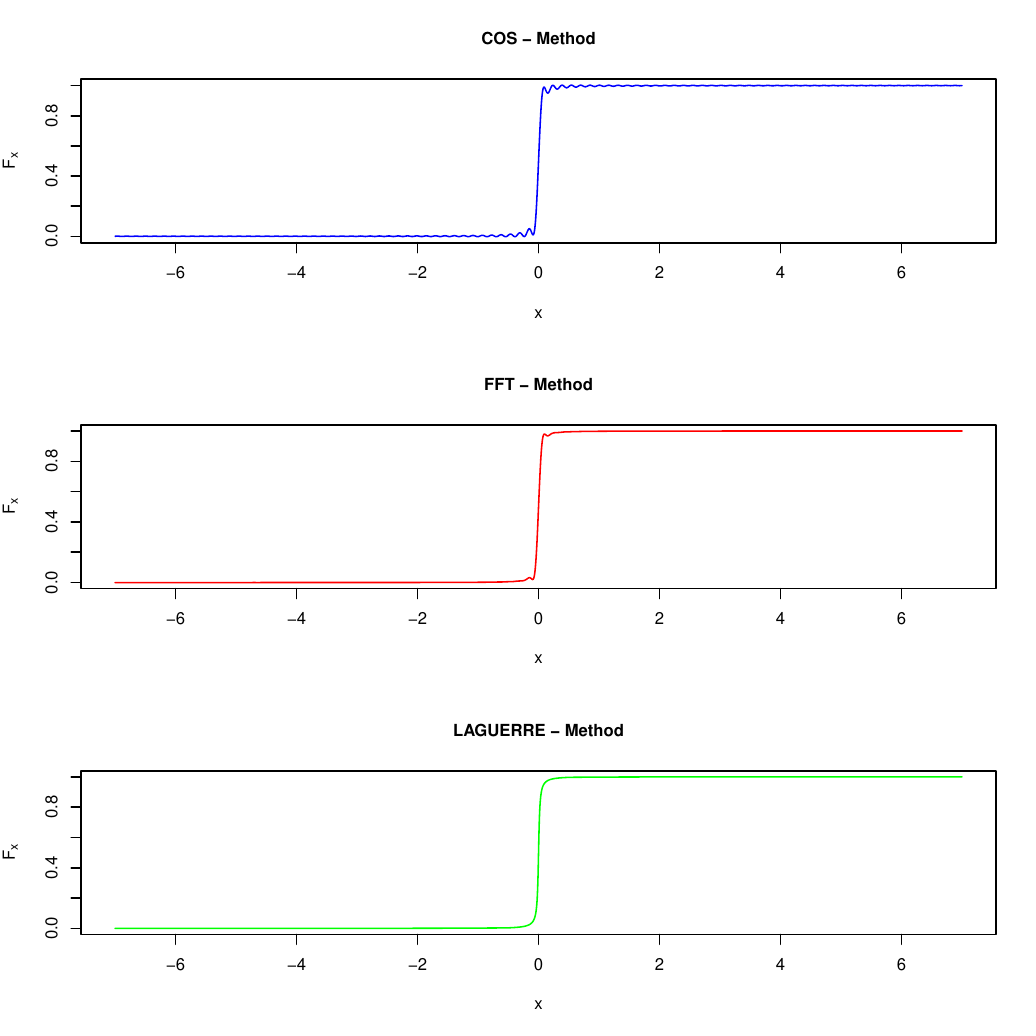}
	\caption{Cumulative distribution function comparison for a Student-$t$ L\'evy increment with $\nu=3$ and $h=0.01$.}
	\label{fig:CDF_Comparison_nu3_h001}
\end{figure}

We conclude this section with a numerical comparison among the three methods implemented in \yuima. To assess the precision of our code we use as a target the cumulative distribution function of a Student-$t$ with $\nu = 3$ computed through the \textsf{R} function \code{pt} available in the package \code{stats}. To conduct this comparison, we construct three \code{yuima.th-objects} as displayed in the code snippet reported below.
\begin{CodeChunk}
\begin{CodeInput}
# To instal YUIMA from Github repository 
R> library(devtools)
R> install_github("yuimaproject/yuima")
##########
# Inputs #
##########
R> library(yuima)
R> nu <- 3
R> h <- 1 # step size for the interval time
R> up <- 10
R> low <- -10

# Support definition for variable x
R> x <- seq(low,up,length.out=100001) 

# Definition of yuima.th-object
R> law_LAG <- setLaw_th(h = h, method = "LAG", up = up, low = low, N = 180) # Laguerre
R> law_COS <- setLaw_th(h = h, method = "COS", up = up, low = low, N = 180) # COS 
R> law_FFT <- setLaw_th(h = h, method = "FFT", up = up, low = low, N = 180) # FFT 

# Cumulative Distribution Function: we apply the cdf method to the yuima.th-object
R> Lag_time <- system.time(ycdf_LAG <- cdf(law_LAG, x, list(nu=nu))) # Laguerre
R> COS_time <- system.time(ycdf_COS<-cdf(law_COS, x, list(nu=nu))) # COS
R> FFT_time <- system.time(ycdf_FFT<-cdf(law_FFT, x, list(nu=nu))) # FFT
\end{CodeInput}
\end{CodeChunk}  
User can specify the numerical methods of the inversion of the characteristic function using the input \code{method}. This argument assumes three values: \code{"LAG"} for the Gauss-Laguerre quadrature, \code{"COS"} for the Cosine Series Expansion and \code{"FFT"} for the Fast Fourier Transform.  Once the \code{yuima.th-object} has been constructed, its cumulative distribution function is computed by applying the \yuima method\footnote{An object of \code{yuima.th-class} inherits the \code{dens} method for the density computation, \code{cdf} for the evaluation of the cumulative distribution function, \code{quantile} for the quantile function and \code{rand} for the generation of the random sample. We refer to \cite{Masuda2022} for the usage of these methods.} \code{cdf} that returns the cumulative distribution function for the \code{numeric} vector \code{x}.  
\begin{figure}[!htbp]
	\centering
		\includegraphics[width=.75\textwidth]{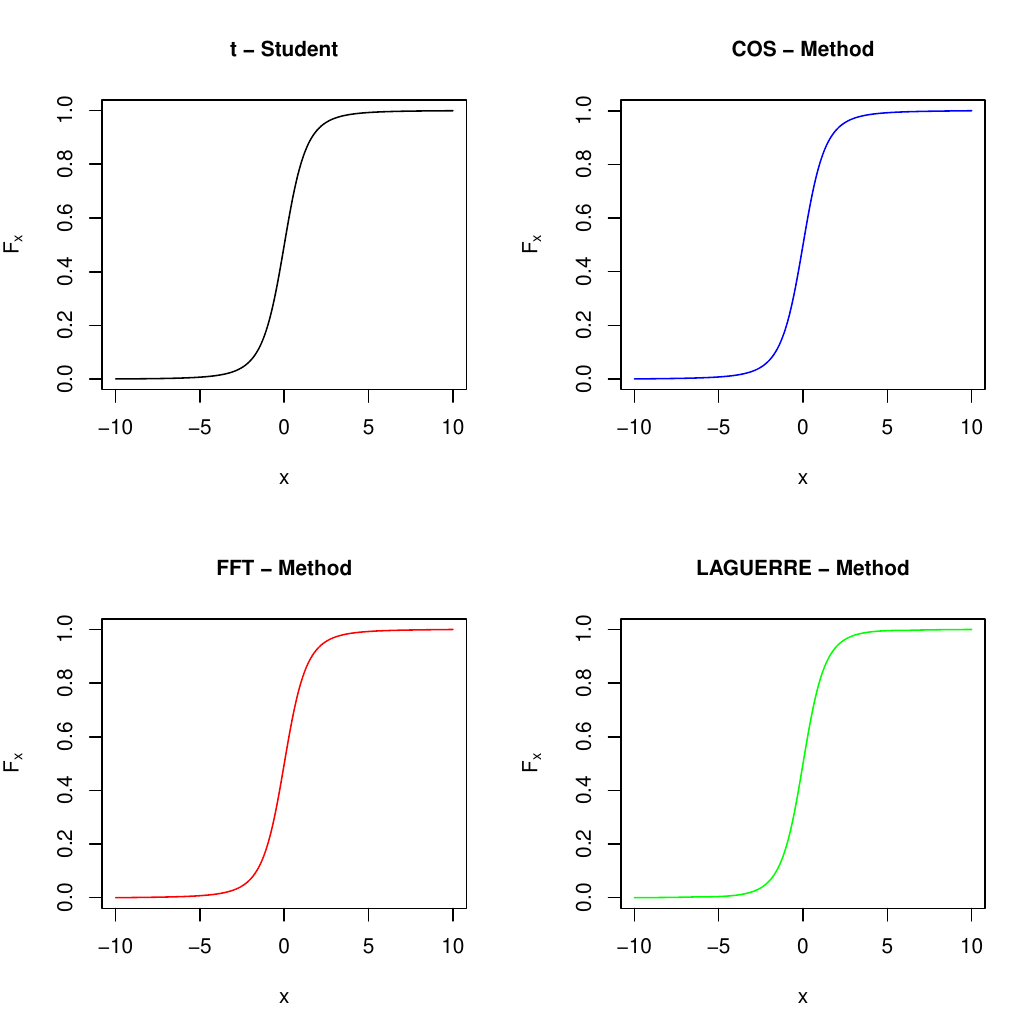}
	\caption{Cumulative distribution function comparison for a Student-$t$ random variable with $\nu=3$.}
	\label{fig:CDF_Comparison_nu3_h1}
\end{figure}
Figure \ref{fig:CDF_Comparison_nu3_h1} compares the cumulative distribution functions of the Student-$t$ obtained using the methods in \yuima with the \code{pt} function. In all cases, we have a good level of precision that is also confirmed by Table \ref{Table2Example1}. As expected the fastest method is the \code{FFT} which seems to be also the most precise. 
\begin{table}[!htbp]
\begin{tabular}{lccc}
\hline\hline
			& \texttt{COS} &  \texttt{FFT} & \texttt{LAG}\\
\hline
RMSE & 0.021 & 0.021 & 0.064\\
Max  & 0.032 & 0.032 & 0.085\\
Min  & 7.88e-05 & 2.18e-05 & 4.97e-04\\
\hline
sec. & 1.47 & 4.00e-02 & 1.2\\
\hline\hline
\end{tabular}
\caption{Summary comparison of the distance between each method available in \yuima and the cumulative distribution function obtained using the \textsf{R} function \code{pt}.}
\label{Table2Example1}
\end{table}
To further investigate this fact, we study the behaviour of the cumulative distribution function when $h$ varies.  For $h= 0.01$, we observe an oscillatory behaviour on tails for the \code{FFT} and \code{COS} while the Laguerre method seems to be more stable as shown in Figure \ref{fig:CDF_Comparison_nu3_h001}. To have a fair comparison, we set \code{N = 180}, however we notice that the precision of \code{FFT} can be drastically improved by tuning the inputs \code{N}, \code{up} and \code{low}.

\section{Numerical examples \label{sec:NE}}

This section presents a series of numerical examples demonstrating the practical application of newly developed classes and methods within the Student-$t$ Regression model. Specifically, we showcase the simulation and estimation of models where the regressors are determined by deterministic functions of time in the first example. The second example introduces integrated stochastic regressors. Lastly, we perform an analysis using real data in the final example.

\subsection{Model with deterministic regressors.}

In this example, we consider two deterministic regressors and the dynamics of the model have the following form:
\begin{equation}
Y_t =\mu_1 \cos\left(5t\right)+\mu_2 \sin\left(t\right) +\sigma J_t
\label{Regr1}
\end{equation}
with the true values $\left(\mu_1, \mu_2, \sigma_0, \nu_0\right)=\left(5,-1,3,3\right)$. \tcb{The estimation of the model \eqref{Regr1} has been investigated in \cite{masuda2023quasilikelihood} where the empirical distribution of the studentized estimates has been discussed}.
To use the simulation method in \yuima we have to write the dynamics of the regressors $X_{1,t}:=\cos\left(5t\right)$ and $X_{2,t}=:\sin\left(5t\right)$, that is:
\begin{equation}
\mbox{d}\left[\begin{array}{c}
X_{1,t}\\
X_{2,t}
\end{array}
\right]=\left[\begin{array}{c}-5\sin\left(5t\right)\\
\cos\left(t\right)\end{array}\right]\mbox{d}t
\label{eq:reg1}
\end{equation}
with the initial condition: 
\[
X_{1,0}=1, \ X_{2,0}=0.
\]
In the following, we show how to implement this model in \yuima.
In this example, we set $h_n=1/50$, and thus the number of the observations $n$ over unit time is $50$. We also set the terminal time of the whole observations $T_n=50$ and that of the partial observations $B_n=15$. 

\begin{CodeChunk}
\begin{CodeInput}
R> library(yuima)
##########
# Inputs #
##########
# Inputs for Fourier Inversion
R> method_Fourier = "FFT"; up = 6; low = -6; N = 2^17; N_grid = 60000 

# Inputs for the sample grid time
R> Factor <- 1
R> Factor1 <- 1
R> initial <- 0; Final_Time <- 50 * Factor; h <- 0.02/Factor1 
\end{CodeInput}
\end{CodeChunk}

We notice that the variable \code{Factor1} controls the step size of the time grid. Indeed for different values of this quantity, we have a different value for $h$ for example  \code{Factor1 = 1, 2, 4, 7.2} corresponds \code{h = 0.02, 0.01, 0.005, 1/365}.

The first step is the definition of an object of \code{yuima.th-class} using the constructor \verb|setLaw_th|.
This object contains the random number generator, the density function, the cumulative distribution function and the quantile function for constructing the increments of a Student-$t$ L\'evy process.

\begin{CodeChunk}
\begin{CodeInput}
#######################################
# Example 1: Deterministic Regressors #
#######################################

R> mu1 <- 5; mu2 <- -1; scale <- 3; nu <- 3 # Model Parameters

# Model Definition
R> law1 <- setLaw_th(method = method_Fourier, up = up, low = low,
  N = N, N_grid = N_grid) # yuima.th 

R> class(yuima_law)
\end{CodeInput}
\begin{CodeOutput}
[1] "yuima.th"
attr(,"package")
[1] "yuima"
\end{CodeOutput}
\end{CodeChunk}

The next step is to define the dynamics of the regressors described in \eqref{eq:reg1}. This set of differential equations is defined in \yuima using the standard constructor \code{setModel}. Once an object containing the mathematical description of the regressors has been defined, we use \code{setLRM} to obtain the \code{yuima.LevyRM}. In the following, we report the command lines for the definition of the Student L\'evy Regression Model in \eqref{Regr1}.

\begin{CodeChunk}
\begin{CodeInput}
R> regr1 <- setModel(drift = c("-5*sin(5*t)", "cos(t)"), diffusion = matrix("0",2,1), 
     solve.variable = c("X1","X2"), xinit = c(1,0)) # Regressors definition

R> Mod1 <- setLRM(unit_Levy = law1, yuima_regressors = regr1) # t-Regression Model

R> class(Mod1)
\end{CodeInput}
\begin{CodeOutput}
[1] "yuima.LevyRM"
attr(,"package")
[1] "yuima"
\end{CodeOutput}
\end{CodeChunk}

Using the object \code{Mod1} we simulate a trajectory of the model in \eqref{Regr1} using the \yuima method \code{simulate}.

\begin{figure}[htbp!]
	\centering
		\includegraphics[width=0.75\textwidth]{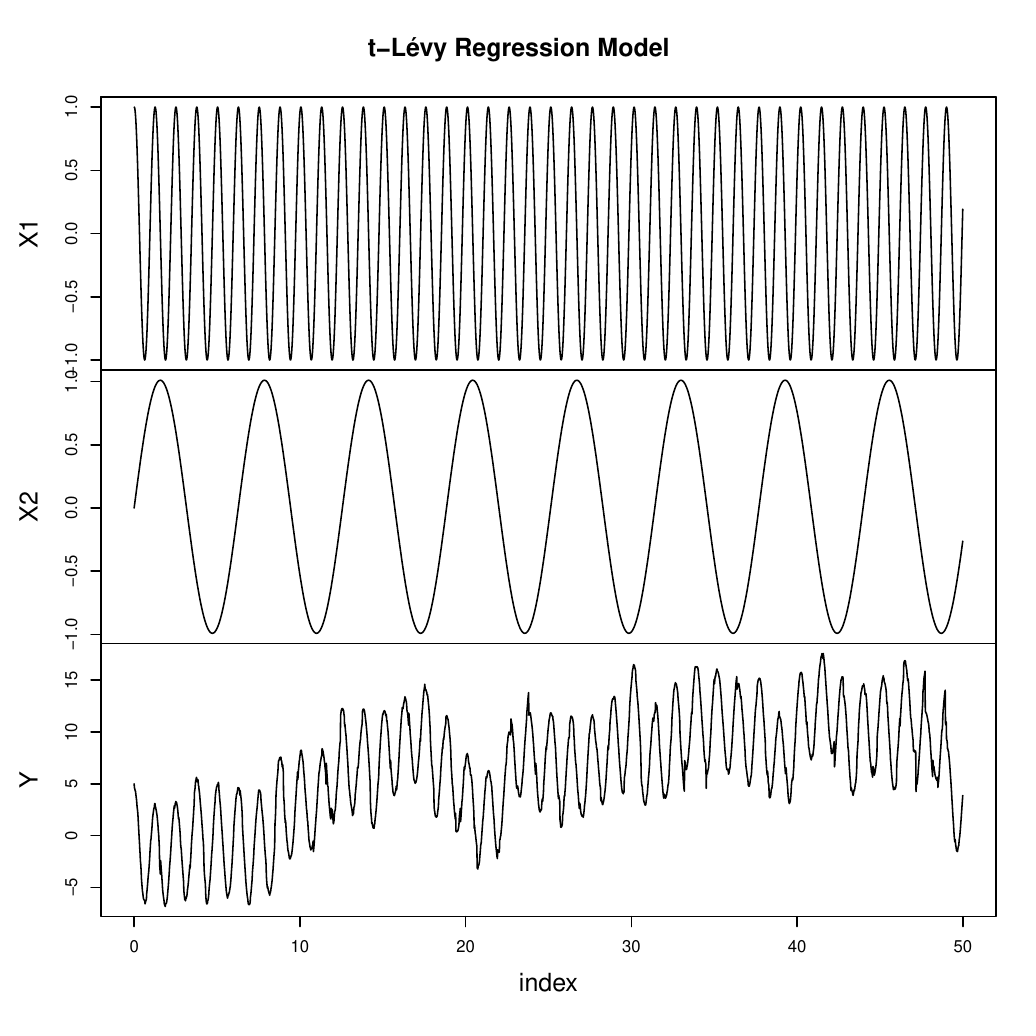}
	\caption{Simulated Trajectory of the Student L\'evy Regression model defined in \eqref{Regr1}. \label{fig:Simulation_Mod1}}
\end{figure}

\begin{CodeChunk}
\begin{CodeInput}
# Simulation
R> samp <- setSampling(initial, Final_Time, n = Final_Time/h)
R> true.par <- unlist(list(mu1 = mu1, mu2 = mu2, sigma0 = scale, nu = nu))

R> set.seed(1)

R> sim1 <- simulate(Mod1, true.parameter = true.par, sampling = samp)
\end{CodeInput}
\end{CodeChunk}
Figure \ref{fig:Simulation_Mod1} reports the simulated sample paths for the regressors $X_{1,t}$, $X_{2,t}$ and the Student L\'evy Regression model $Y_{t}$.
The next step is to study the behaviour of the two-step estimation procedure described in Section \ref{Math:Model}. To run this procedure, we use the method \verb|estimation_LRM| and then we initialize randomly the optimization routine as shown in the following command lines.

\begin{CodeChunk}
\begin{CodeInput}
# Estimation
R> lower <- list(mu1 = -10, mu2 = -10,  sigma0 = 0.1)
R> upper <- list(mu1 = 10, mu2 = 10, sigma0 = 10.01)
 
R> start <- list(mu1 = runif(1, -10, 10), mu2 = runif(1, -10, 10), 
+    sigma0 = runif(1, 0.01, 4))

R> Bn <- 15*Factor

R> est1 <- estimation_LRM(start = start, model = sim1, data = sim1@data,
+    upper = upper, lower = lower, PT = Bn)

R> class(est1)
\end{CodeInput}
\begin{CodeOutput}
[1] "yuima.qmle"
attr(,"package")
[1] "yuima"
\end{CodeOutput}
\end{CodeChunk}

\begin{CodeChunk}
\begin{CodeInput}
R> summary(est1)
\end{CodeInput}
\begin{CodeOutput}
Quasi-Maximum likelihood estimation

Call:
estimation_LRM(start = start, model = sim1, data = sim1@data, 
    upper = upper, lower = lower, PT = Bn)

Coefficients:
        Estimate Std. Error
mu1     5.029555 0.03538355
mu2    -1.128905 0.18026088
sigma0  2.425147 0.12523404
nu      2.735344 0.47457435

-2 log L: 3236.535 73.3833 
\end{CodeOutput}
\end{CodeChunk}

It is also possible to construct the dataset without the method \code{simulate}. This result can be achieved by constructing an object of \code{yuima.th-class} that represents the increment of the Student-$t$ L\'evy process on the time interval with length $h$. 

\begin{CodeChunk}
\begin{CodeInput}
# Dataset construction without YUIMA
R> time_t <- index(get.zoo.data(sim1@data)[[1]])
R> X1 <- zoo(cos(5*time_t), order.by = time_t)
R> X2 <- zoo(sin(time_t), order.by = time_t)
R> law1_h <- setLaw_th(h = h, method = method_Fourier, up = up, low = low,
     N = N, N_grid = N_grid) 

R> print(c(law1_h@h,law1@h))
\end{CodeInput}
\begin{CodeOutput}
[1] 0.02 1.00
\end{CodeOutput}
\end{CodeChunk}
The object \verb|law1_h| refers to the Student-$t$ L\'evy increment over an interval with length $0.02$ as it can be seen looking at the slot \code{...@h}. 
\begin{CodeChunk}
\begin{CodeInput}
R> set.seed(1)
R> names(nu)<- "nu"
R> J_t <- zoo(cumsum(c(0,rand(law1_h, Final_Time/h,nu))), order.by = time_t)
R> Y <- mu1 * X1 + mu2 * X2 + scale/sqrt(nu) * J_t
R> data1_a <- merge(X1, X2, Y)
\end{CodeInput}
\end{CodeChunk}

The estimation is performed by means of the method \verb|estimation_LRM| as in the previous example.
\begin{CodeChunk}
\begin{CodeInput}
R> est1_a <- estimation_LRM(start = start, model = Mod1, data = setData(data1_a),
+    upper = upper, lower = lower, PT = Bn)
R> summary(est1_a)
\end{CodeInput}
\begin{CodeOutput}
Quasi-Maximum likelihood estimation

Call:
estimation_LRM(start = start, model = Mod1, data = setData(data1_a), 
    upper = upper, lower = lower, PT = Bn)

Coefficients:
        Estimate Std. Error
mu1     4.987778 0.03642823
mu2    -1.164138 0.18558422
sigma0  2.495930 0.12888927
nu      2.407683 0.41221590

-2 log L: 3232.784 85.32438 
\end{CodeOutput}
\end{CodeChunk}

Using the result stored in the \code{summary} we can construct a confidence interval using the command lines reported below.

\begin{CodeChunk}
\begin{CodeInput}
R> info_sum <- summary(est1_a)@coef
R> alpha <- 0.025
R> Confidence_Int_95 <- rbind(info_sum[,1]+info_sum[,2]*qnorm(alpha),
+     info_sum[,1]+info_sum[,2]*qnorm(1-alpha),unlist(true.par),
+     info_sum[,1])
R> rownames(Confidence_Int_95) <- c("LB","UB","True_par","Est_par")
R> print(Confidence_Int_95, digit = 4)
\end{CodeInput}
\begin{CodeOutput}
           mu1      mu2  sigma0     nu
LB       4.916  -1.5279   2.243  1.600
UB       5.059  -0.8004   2.749  3.216
True_par 5.000  -1.0000   3.000  3.000
Est_par  4.988  -1.1641   2.496  2.408
\end{CodeOutput}
\end{CodeChunk}
\color{black}
Based on the aforementioned examples, the estimated value for the $\sigma$ parameter deviates from its true value. To further investigate this observation, we conducted a comparative analysis using the three integration methods discussed in Section \ref{yuimath}. This exercise was repeated for three different values of $h= 0.01, 0.005$ and $\frac{1}{365}$. To ensure a fair comparison among the Laguerre, Cosine Series expansion, and Fast Fourier Transform methods, we set the argument \code{N = 180}. The obtained results are presented in Table \ref{Tab:CompN180}.


\begin{table}[!htbp]
\begin{tabular}{lccccccccc}
\hline\hline
& \multicolumn{3}{c}{Laguerre Method} & \multicolumn{3}{c}{COS Method} & \multicolumn{3}{c}{FFT Method}\\
\hline 
 $h$       &    0.01 & 0.005   & $1/365$ & 0.01    & 0.005   & $1/365$ &  0.01   &   0.005   & $1/365$ \\ 
\hline
$\hat{\mu_1}$    &  4.968  & 5.019   & 4.976   & 4.939   & 5.065   &  4.875  & 4.934   &  5.069    & 4.876\\
           &  \footnotesize{(0.029)}& \footnotesize{(0.020)} & \footnotesize{(0.014)} & \footnotesize{(0.041)} & \footnotesize{(0.049)} & \footnotesize{(0.054)} & \footnotesize{(0.042)} &  \footnotesize{(0.049)}  & \footnotesize{(0.054)}\\
$\hat{\mu_2}$    & -1.065  & -1.044  & -0.914  & -1.192  & -1.195  & -0.555  & -1.179  &  -1.178   & -0.528\\
           &  \footnotesize{(0.146)}& \footnotesize{(0.104)} & \footnotesize{(0.074)} & \footnotesize{(0.210)}  & \footnotesize{(0.247)} & \footnotesize{(0.277)} & \footnotesize{(0.213)} &  \footnotesize{(0.248)}  & \footnotesize{(0.277)}\\
$\hat{\sigma_0}$ &  2.770  & 2.794   & 2.657   & 3.993   & 6.654   & 10.010  & 4.055   &  6.674    & 10.010\\
           &  \footnotesize{(0.101)}& \footnotesize{(0.072)} & \footnotesize{(0.051)} & \footnotesize{(0.145)} & \footnotesize{(0.172)} & \footnotesize{(0.192)} & \footnotesize{(0.148)} &  \footnotesize{(0.172)}  & \footnotesize{(0.193)}\\
$\hat{\nu}$      &  3.462  & 3.281   & 3.067   & 2.827   & 2.223   &  2.227  & 5.749   &  8.310    & 15.175\\
           &  \footnotesize{(0.615)}& \footnotesize{(0.580)} & \footnotesize{(0.538)} & \footnotesize{(0.492)} & \footnotesize{(0.377)} & \footnotesize{(0.378)}& \footnotesize{(1.063)} &  \footnotesize{(1.571)}  & \footnotesize{(2.940)}\\
\hline
sec. & 2.07 & 2.02 & 2.07 & 1.10 & 1.24 & 1.34 & 0.15 & 0.24 & 0.39\\
\hline\hline
\end{tabular}
\caption{Estimated parameters for $h = 0.01, 0.005, 1/365$ and different integration methods. The number of points for the inversion of the characteristic function is $N=180$\label{Tab:CompN180}. The parenthesis shows the asymptotic standard error. The last row reports the seconds necessary for the simulation of a sample path. }
\end{table}

In the Laguerre method, we observe that reducing the step size $h$ leads to improved estimates of $\nu$, however looking at the asymptotic standard error, the estimates for $\sigma_0$ seem to maintain the bias. As for the remaining methods, the estimates exhibit unsatisfactory performance due to the imposed restriction of \code{N = 180} (that is the maximum value allowed for the Laguerre quadrature in the evaluation of the density in \eqref{eq:InvForLM}). This outcome is not unexpected, given the Laguerre method's ability to yield a valid distribution even with a relatively small value of \code{N}, as demonstrated in Figure \ref{fig:CDF_Comparison_nu3_h001}. Nevertheless for the \code{COS} and the \code{FFT}, it is possible to enhance the results by increasing the value of the argument \code{N} that yields a more accurate result in the simulation of the sample path for the model described in \eqref{Regr1}.

\begin{table}[!htbp]
\begin{tabular}{lcccccc}
\hline\hline
           & \multicolumn{3}{c}{COS Method} & \multicolumn{3}{c}{FFT Method}\\
\hline
\code{N}   &  1000    & 5000    & 10000   & 1000    & 5000    & 10000  \\
$\mu_1$    &  4.968   & 4.975   & 4.975   & 4.968   & 4.975   & 4.975  \\
           &  \footnotesize{(0.034)} & \footnotesize{(0.016)} & \footnotesize{(0.016)} & \footnotesize{(0.035)} & \footnotesize{(0.306)} & \footnotesize{(0.016)} \\
$\mu_2$    &  -0.887  & -0.909  & -0.909  & -0.886  & -0.908  & -0.908\\
           &  \footnotesize{(0.174)} & \footnotesize{(0.082)} & \footnotesize{(0.082)} & \footnotesize{(0.177)} & \footnotesize{(0.022)} & \footnotesize{(0.082)}\\
$\sigma_0$ &  3.300   & 2.964   & 2.964   & 3.363   & 2.963   & 2.964\\
           &  \footnotesize{(0.064)} & \footnotesize{(0.057)} & \footnotesize{(0.057)} & \footnotesize{(0.065)} & \footnotesize{(0.057)} & \footnotesize{(0.057)}\\
$\nu$      &  2.713   & 2.765   & 2.765   & 3.279   & 2.760   & 2.702\\
           &  \footnotesize{(0.470)} & \footnotesize{(0.480)} & \footnotesize{(0.480)} & \footnotesize{(0.579)} & \footnotesize{(0.479)} & \footnotesize{(0.468)}\\
\hline
sec.       &    3.49  & 14.83   & 33.70   & 0.38  &  0.44 & 0.45\\
\hline\hline
\end{tabular}
\caption{Estimated parameters for $N = 1000, 5000, 10^4$ in the \code{COS} and the \code{FFT} methods. The step size $h$ is $1/365$. The parenthesis shows the asymptotic standard error. The last row reports the seconds necessary for the simulation of a sample path. \label{Tab:Comph365} }
\end{table}

Table \ref{Tab:Comph365} shows the estimated parameters for the varying value of \code{N} and $h=1/365$. As expected increasing the precision in the quadrature improved estimates for both methods. Notably for \code{N} $\geq$ 5000, all estimates fall within the asymptotic confidence interval at the $95\%$ level.

%


\subsection{Model with integrated stochastic regressors}

In this section, we consider two examples whose regressors are stochastic.
Moreover, to satisfy the regularity conditions, the regressors are supposed to be an integrated version of stochastic processes.

\paragraph{Example 1} 
In this example, we consider the following continuous time regression model with a single regressor: 
\begin{equation}\label{yu:ex1}
 Y_t=\mu \int_0^t X_{u}du+\sig J_t, \quad J_1\sim t_\nu,
\end{equation}
with the true values $(\mu_0,\sig_0,\nu_0)=(-3,3,2.5)$, and the process $X$ is supposed to be the L\'{e}vy driven Ornstein-Uhlenbeck process defined as:
\begin{equation}
\nn dX_t= -X_tdt+ 2dZ_t,
\end{equation}
where the driving noise $Z$ is the L\'{e}vy process with $Z_1\sim\text{NIG}(1,0,1,0)$.
The normal inverse Gaussian (NIG) random variable is defined as the normal-mean variance mixture of the inverse Gaussian random variable, and the probability density function of $Z_t\sim \text{NIG}(\al,\beta,\del t,\mu t)$ is given by
\begin{equation*}
x\mapsto
\frac{\alpha\delta t\exp\{\delta t\sqrt{\alpha^{2}-\beta^{2}}+\beta(x-\mu t)\} K_{1}(\alpha \psi(x;\delta t,\mu t))}{\pi \psi(x;\delta t,\mu t)}
\end{equation*}
where $\alpha^2:=\gamma^2+\beta^2$ and $\psi(x;\delta t,\mu t):=\sqrt{(\delta t)^{2}+(x-\mu t)^{2}}$.
More detailed theoretical properties of the NIG-L\'{e}vy process are given for example in \cite{Bar98}.

For the simulation by \yuima, we formally introduce the following system:
\begin{equation}
d\Bigg[\begin{array}{c} X_{1,t}\\ X_{2,t}\end{array}\Bigg]=\Bigg[\begin{array}{c} X_{2,t}\\ -X_{2,t}\end{array}\Bigg]dt+\Bigg[\begin{array}{c} 0\\ 2\end{array}\Bigg]dZ_t,
\end{equation}
with the initial condition: 
\[
X_{1,0}=0, \ X_{2,0}=0.
\]
The process $X_{1,t}$ corresponds to the \tcb{regressor $\int_0^{t}X_{u}du$ in the regression model \eqref{yu:ex1}}.
Due to the specification of the new \yuima function, we will additionally construct the ``full" model:
\begin{equation}
Y_t=\mu_1 X_{1,t}+\mu_2 X_{2,t}+\sig J_t, \quad J_1\sim t_\nu,
\label{yu:ex1f}
\end{equation}
and simulate the trajectory of $Y$ with $\mu_2=0$.
After that, we will estimate the parameters $(\mu_1,\sig,\nu)$ by the original model \eqref{yu:ex1}.
From now on, we show the implementation of this model in \yuima.
The sampling setting is unchanged from the previous example code:
$h_n=1/50,  n=1/h_n=50,  T_n=50,  B_n=15$.

\begin{CodeChunk}
\begin{CodeInput}
R> library(yuima)
##########
# Inputs #
##########
# Inputs for Fourier Inversion
R> method_Fourier = "FFT"; up = 6; low = -6; N = 2^17; N_grid = 60000 

# Inputs for the sample grid time
R> Factor <- 1
R> Factor1 <- 1
R> initial <- 0; Final_Time <- 50 * Factor; h <- 0.02/Factor1 
\end{CodeInput}
\end{CodeChunk} 

The role of the variable \code{Factor1} is the same as in the previous section.
By using the constructor \verb|setLaw_th|, we define \code{yuima.th-class} in a similar manner. 
After that, we construct the ``full" model \eqref{yu:ex1f} by the constructors \code{setModel} and \code{setLRM}.
A trajectory of $Y$ and its regressors can be simulated by the YUIMA method \code{simulate}.

\begin{CodeChunk}
\begin{CodeInput}
####################################################
# Regressor: Integrated NIG-Levy driven OU process #
####################################################
R> mu1 <- -3; mu2 <- 0;  scale <- 3; nu <- 2.5 # Model Parameters

# Model Definition
R> lawILOU <- setLaw_th(method = method_Fourier, up = up, low = low, N = N, 
+    N_grid = N_grid) # yuima.th

R> regrILOU <- setModel(drift = c("X2", "-X2"), jump.coeff = c("0", "2"), 
+    solve.variable = c("X1","X2"), xinit = c(0,0), measure.type = "code",
+    measure = list(df = "rNIG(z, 1, 0, 1, 0)")) # Regressors definition

# t-Regression model
R> ModILOU <- setLRM(unit_Levy = lawILOU, yuima_regressors = regrILOU) 

# Simulation
R> sampILOU <- setSampling(initial, Final_Time, n = Final_Time/h)
R> true.parILOU <- unlist(list(mu1 = mu1, mu2 = mu2, sigma0 = scale, nu = nu))
R> set.seed(1)
R> simILOU <- simulate(ModILOU, true.parameter = true.parILOU, sampling = sampILOU)
\end{CodeInput}
\end{CodeChunk}   

\begin{figure}[htbp!]
	\centering
		\includegraphics[width=0.75\textwidth]{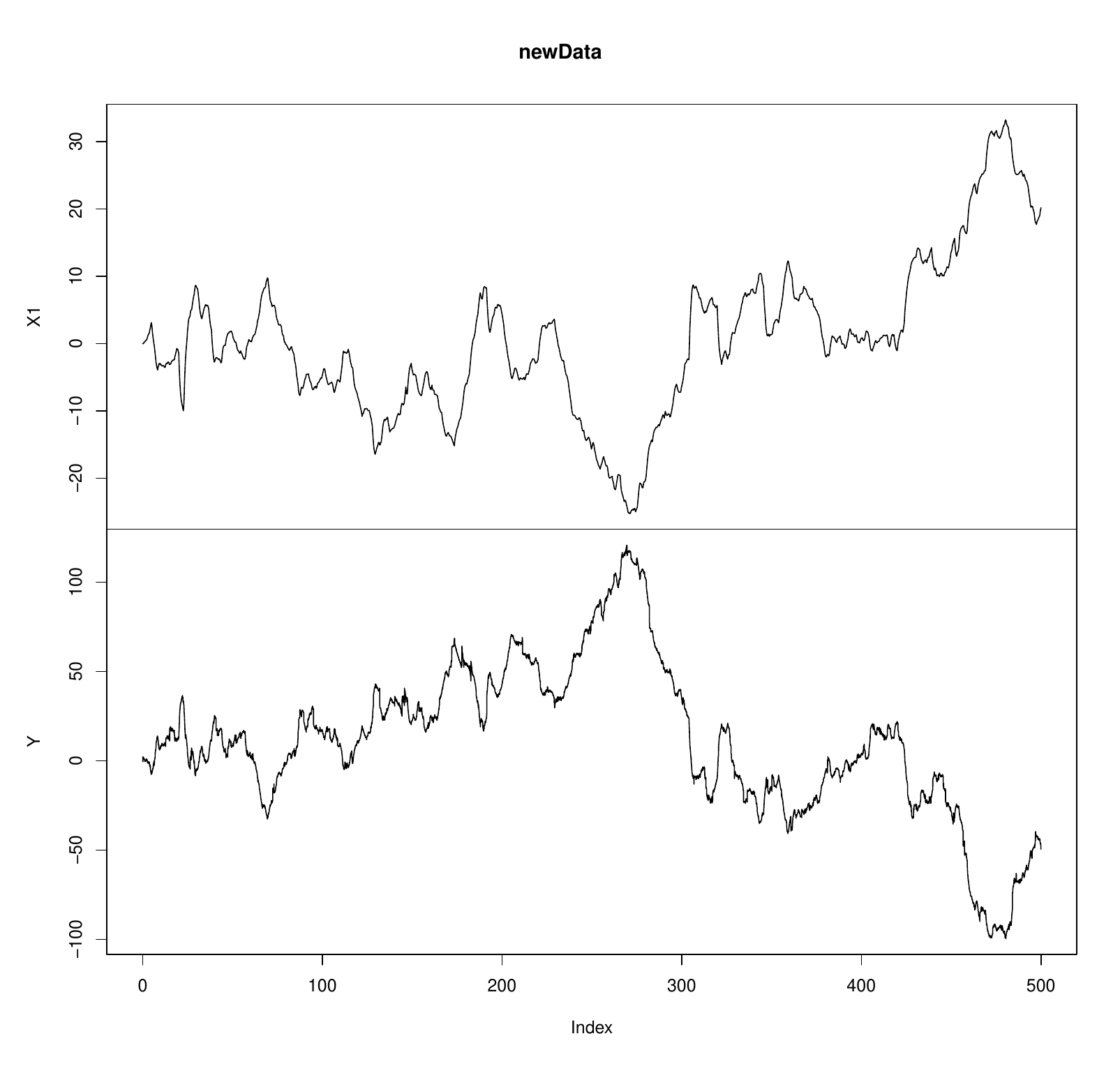}
	\caption{Simulated Trajectory of the Student L\'evy Regression model defined in \eqref{yu:ex1}. \label{fig:Simulation_Mod2}}
\end{figure}

Next we extract the trajectories of $Y$ and $X_{1}$ from the \code{yuima.LevyRM-class}: \code{simILOU} and construct the new model for the estimation of the parameters.
Figure \ref{fig:Simulation_Mod2} shows the simulated sample paths for the regressor and \tcb{the response process} $Y$.

\begin{CodeChunk}
\begin{CodeInput}
# Data extraction
R> Dataset<-get.zoo.data(simILOU)
R> newData <-Dataset[-2]
R> newData <- merge(newData$X1, newData$Y)
R> colnames(newData) <- c("X1","Y")
R> plot(newData)

# Define the model for estimation
R> regrILOU1 <- setModel(drift = c("0"), diffusion = matrix(c("0"),1,1), 
+    solve.variable = c("X1"), xinit = c(0)) # Regressors definition

# t-Regression model
R> ModILOU1 <- setLRM(unit_Levy = lawILOU, yuima_regressors = regrILOU1) 

# Estimation
R> lower1 <- list(mu1 = -10,  sigma0 = 0.1)
R> upper1 <- list(mu1 = 10, sigma0 = 10.01)
R> startILOU1<- list(mu1 = runif(1, -10, 10), sigma0 = runif(1, 0.01, 4))

R> Bn <- 15*Factor
R> Data1 <- setData(newData)
R> estILOU1 <- estimation_LRM(start = startILOU1, model = ModILOU1, data = Data1,
+    upper = upper1, lower = lower1, PT = Bn)
R> summary(estILOU1)
\end{CodeInput}
\begin{CodeOutput}
Quasi-Maximum likelihood estimation

Call:
estimation_LRM(start = startILOU1, model = ModILOU1, data = Data1, 
    upper = upper1, lower = lower1, PT = Bn)

Coefficients:
        Estimate Std. Error
mu1    -2.959621 0.02112734
sigma0  2.778659 0.03208519
nu      2.404843 0.13018410

-2 log L: 69711.32 854.3838 
\end{CodeOutput}
\end{CodeChunk}

\paragraph{Example 2}

In this example, we consider the following regression model:
\begin{equation}\label{yu:ex2}
Y_t= \mu_1\int_0^t V_{1,u}du+\mu_2\int_0^t V_{2,u}du+\sig J_t, \ \quad J_1\sim t_\nu\end{equation}
where the process $X=(V_{1},V_{2})$ satisfy
\begin{align}
&dV_{1,t}=\frac{1}{\epsilon}(V_{1,t}-V_{1,t}^3-V_{2,t}+s)dt,\\
&dV_{2,t}=(\gam V_{1,t}-V_{2,t}+\alpha)dt+\sigma' dw_t,
\end{align}
with 
\[(\ep, s, \gam, \al,\sig')=(1/3, 0, 3/2, 1/2,2),\]
and standard Wiener process $w$.
We set the true values 
\[(\mu_{1,0},\mu_{2,0},\sig_0,\nu_0)=(8,-4,8,3).\]
The process $V$ is the so-called stochastic FitzHugh-Nagumo process which is a classical model for describing a neuron; $V_{1}$ expresses the membrane potential of the neuron and $V_{2}$ represents a recovery variable.
Its theoretical properties such as hypoellipticity, and Feller and mixing properties are well summarized in \cite{LeoSam18}.
The paper also provides a nonparametric estimator of the invariant density and spike rate.
Similarly, other integrated degenerate diffusion can be considered as regressors.
Its ergodicity for the regularity condtions is studied for example in \cite{Wu01}.
For the statistical inference for degenerate diffusion processes, we refer to \cite{DitSam19} and \cite{GloYos21}.

\tcb{For the implementation of the regression model \eqref{yu:ex2} on YUIMA, we formally consider the following dynamics:
\begin{equation}
d\left[\begin{array}{c} V_{1,t}\\ V_{2,t}\\ V_{3,t}\\ V_{4,t}\end{array}\right]=\left[\begin{array}{c} V_{3,t}\\ V_{4,t}\\ 3(V_{3,t}-V_{3,t}^3-V_{4,t})\\ \frac{3}{2}V_{3,t}-V_{4,t}+\frac{1}{2}\end{array}\right]dt+\left[\begin{array}{c} 0\\ 0\\0\\2\end{array}\right]dw_t,
\end{equation}
with the initial condition:
\[
V_{1,0}=0, \ V_{2,0}=0, \ V_{3,0}=0, \ V_{4,0}=0.
\]
The first and second elements correspond to the regressor in \eqref{yu:ex2}.
As in the previous example, we first simulate data by the ``full" model defined as:
\begin{equation}
Y_t=\mu_1 V_{1,t}+\mu_2V_{2,t}+\mu_3V_{3,t}+\mu_4V_{4,t}+\sig J_t,\quad J_1\sim t_\nu,
\end{equation}
with $\mu_3=\mu_4=0$,} and after that, we extract the simulated data and estimate the parameters based on the original regression model \eqref{yu:ex2}.
Below we show how to implement on YUIMA. 
In this example, we set $h_n=1/200, n=1/h_n=200, T_n=1000, B_n=300$.
The values of them are controlled by the variables \texttt{Factor} and \texttt{Factor1} in the example code.

\begin{CodeChunk}
\begin{CodeInput}
R> library(yuima)
##########
# Inputs #
##########
# Inputs for Fourier Inversion
R> method_Fourier = "FFT"; up = 6; low = -6; N = 2^17; N_grid = 60000 

# Inputs for the sample grid time
R> Factor <- 20
R> Factor1 <- 4
R> initial <- 0; Final_Time <- 50 * Factor; h <- 0.02/Factor1 

############################################################
# Regressor: Integrated stochastic FitzHugh-Nagumo process #
############################################################

R> mu1 <- 8; mu2 <- -4; mu3 <- 0; mu4 <- 0;  scale <- 8; nu <- 3 # Model Parameters
# Model Definition
R> lawFN <- setLaw_th(method = method_Fourier, up = up, low = low,
+    N = N, N_grid = N_grid) # yuima.th 

R> regrFN <- setModel(drift = c("V3", "V4", "3*(V3-V3^3-V4)", "1.5*V3-V4+0.5"), 
+    diffusion = matrix(c("0", "0", "0", "2"),4,1), 
+    solve.variable = c("V1","V2", "V3", "V4"),
+    xinit = c(0,0,0,0)) # Regressors definition

R> ModFN <- setLRM(unit_Levy = lawFN, yuima_regressors = regrFN) #  t-regression model

# Simulation
R> sampFN <- setSampling(initial, Final_Time, n = Final_Time/h)
R> true.parFN <- unlist(list(mu1 = mu1, mu2 = mu2, mu3 = mu3, mu4 = mu4, 
+    sigma0 = scale, nu = nu))
R> set.seed(12)
R> simFN <- simulate(ModFN, true.parameter = true.parFN, sampling = sampFN)
\end{CodeInput}
\end{CodeChunk}   

\begin{figure}[htbp!]
	\centering
		\includegraphics[width=0.75\textwidth]{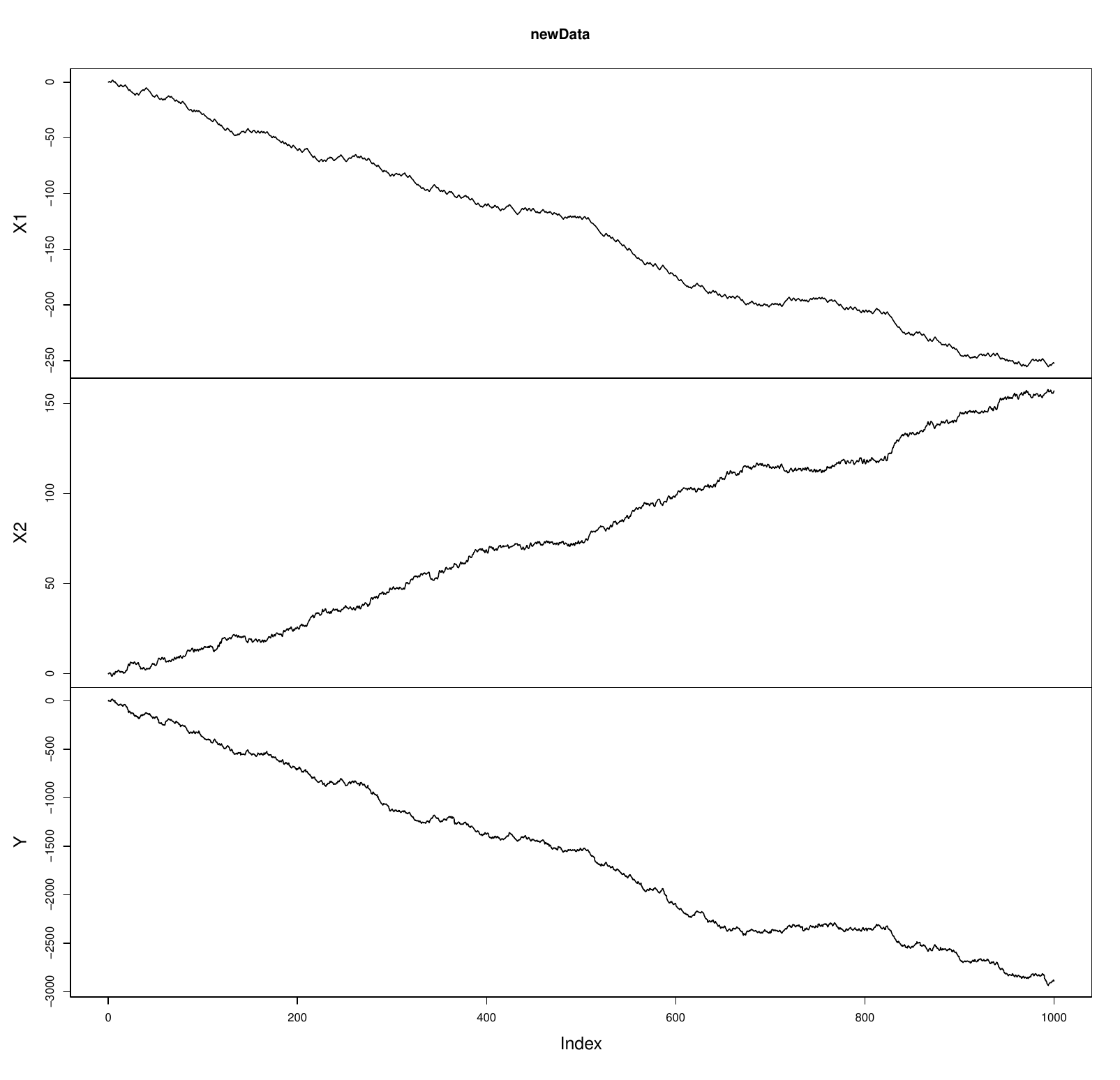}
	\caption{Simulated Trajectory of the Student L\'evy Regression model defined in \eqref{yu:ex2}. \label{fig:Simulation_Mod3}}
\end{figure}

In this example, both \code{Factor} and \code{Factor1} are larger than the previous example, and hence it takes a higher computational load than the previous one.  
Figure \ref{fig:Simulation_Mod3} shows the simulated sample paths for the regressor and the Student L\'{e}vy Regression model $Y$.
Now we move on to the estimation phase.

\begin{CodeChunk}
\begin{CodeInput}
R> Dataset<-get.zoo.data(simFN)
R> newData <-Dataset[-c(3,4)]
R> newData <- merge(newData$X1,newData$X2, newData$Y)
R> colnames(newData) <- c("X1","X2", "Y")
R> plot(newData)

# Define the model for estimation
R> regrFN1 <- setModel(drift = c("0", "0"), diffusion = matrix(c("0", "0"),2,1),
+    solve.variable = c("X1", "X2"), xinit = c(0,0)) # Regressors definition
R> ModFN1 <- setLRM(unit_Levy = lawFN, yuima_regressors = regrFN1) # t-Regression model.

# Estimation
R> lower1 <- list(mu1 = -10, mu2 = -10,  sigma0 = 0.1)
R> upper1 <- list(mu1 = 10, mu2 =10, sigma0 = 10.01)
R> startFN1<- list(mu1 = runif(1, -10, 10), mu2 = runif(1, -10, 10), 
+    sigma0 = runif(1, 0.01, 4))

R> Bn <- 15*Factor
R> Data1 <- setData(newData)

R> estFN1 <- estimation_LRM(start = startFN1, model = ModFN1, data = Data1,
+    upper = upper1, lower = lower1, PT = Bn)
R> summary(estFN1)
\end{CodeInput}
\begin{CodeOutput}
Quasi-Maximum likelihood estimation

Call:
estimation_LRM(start = startFN1, model = ModFN1, data = Data1, 
    upper = upper1, lower = lower1, PT = Bn)

Coefficients:
        Estimate Std. Error
mu1     8.041838 0.04895791
mu2    -4.041518 0.04495255
sigma0  7.712157 0.04452616
nu      3.020972 0.11837331

-2 log L: 403960.9 1291.516 
\end{CodeOutput}
\end{CodeChunk}

\subsection{Real data regressors}
\begin{figure}[!h]
	\centering
		\includegraphics[width=0.75\textwidth]{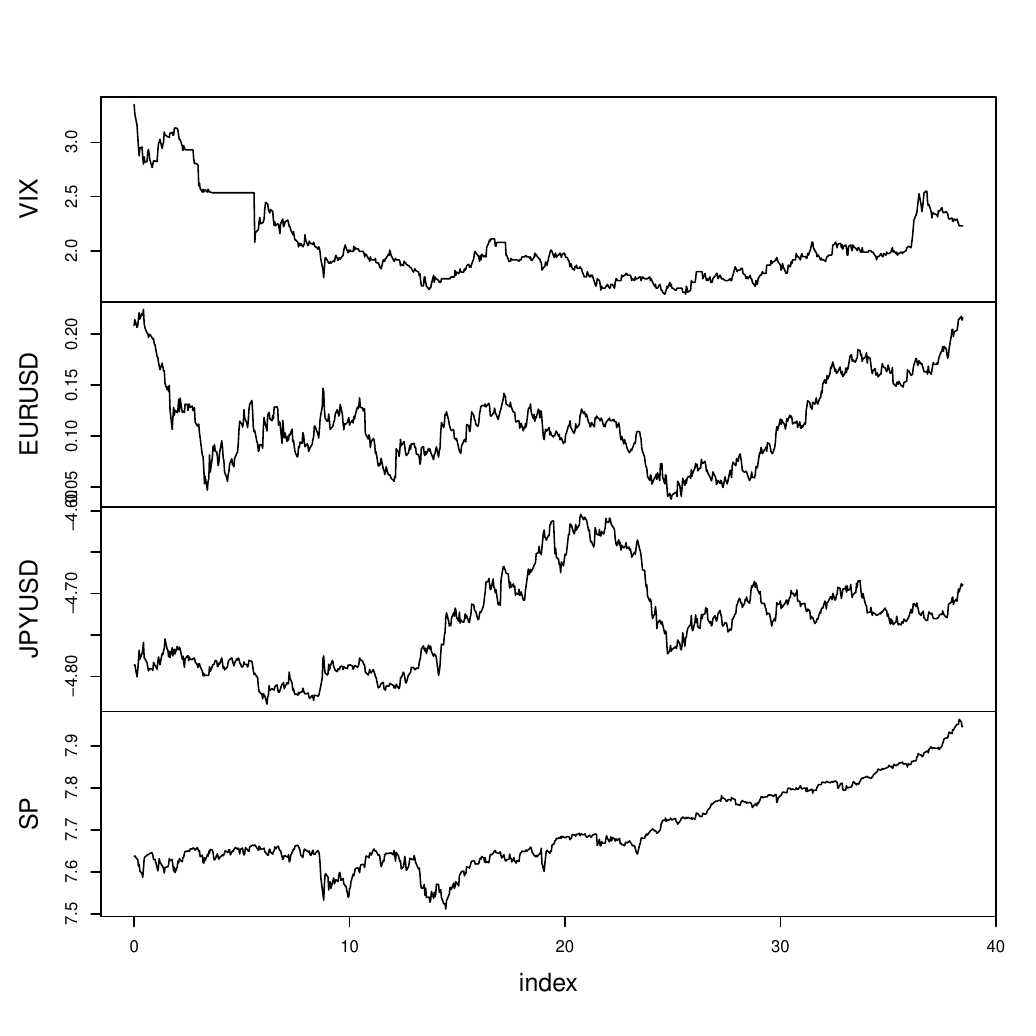}
	\caption{Financial data observed on a monthly equally grid time ranging from December 14th 2014 to May 12th 2023.\label{fig:RealDataset}}
\end{figure}

In this section, we show how to use the \yuima package for the estimation of a Student L\'evy Regression model in a real dataset. We consider a model where the daily price of the Standard and Poor 500 Index is explained by the VIX index and two currency rates: the YEN-USD and the Euro Usd rates. The dataset was provided by \code{Yahoo.finance} and ranges from
December 14th 2014 to May 12th 2023. We downloaded the data using the function \code{getSymbol} available in \code{quantmod} library. To consider a small value for the step size $h$ we estimate our model every month ($h=1/30$). We report below the code for storing the market data in an object of \code{yuima.data-class}
\begin{CodeChunk}
\begin{CodeInput}
R> library(yuima)
R> library(quantmod)
R> getSymbols("^SPX", from = "2014-12-04", to = "2023-05-12")
# Regressors
R> getSymbols("EURUSD=X", from = "2014-12-04", to = "2023-05-12")
R> getSymbols("VIX", from = "2014-12-04", to = "2023-05-12")
R> getSymbols("JPYUSD=X", from = "2014-12-04", to = "2023-05-12")

R> SP <- zoo(x = SPX$SPX.Close, order.by = index(SPX$SPX.Close))
R> Vix <- zoo(x = VIX$VIX.Close/1000, order.by = index(VIX$VIX.Close))
R> EURUSD <- zoo(x = `EURUSD=X`[,4], order.by = index(`EURUSD=X`))
R> JPYUSD <- zoo(x = `JPYUSD=X`[,4], order.by = index(`JPYUSD=X`))
R> Data <- na.omit(na.approx(merge(Vix, EURUSD, JPYUSD, SP)))
R> colnames(Data) <- c("VIX", "EURUSD", "JPYUSD", "SP")
R> days <- as.numeric(index(Data))-as.numeric(index(Data))[1]

# equally spaced grid time data
R> Data <- zoo(log(Data), order.by = days)
R> Data_eq <- na.approx(Data, xout = days[1] : tail(days,1L))
R> yData <- setData(zoo(Data_eq, order.by = index(Data_eq)/30)) # Data on monthly basis
\end{CodeInput}
\end{CodeChunk}
We decided to work with log-price (see \code{Data} variable in the above code) to have quantities defined on the same support of a Student-$t$ L\'evy process, i.e.: the real line. Since the estimation method in \yuima requires that the data are observed on an equally spaced grid time, we interpolate linearly to estimate possible missing data to get a log price for each day. Figure \ref{fig:RealDataset} shows the trajectory for each financial series.
To estimate the model we perform the same steps discussed in the previous examples and we report below the code for reproducing our result.
\begin{CodeChunk}
\begin{CodeInput}
# Inputs for integration in the inversion formula
R> method_Fourier <- "FFT"; up <- 6; low <- -6; N <- 10000; N_grid <- 60000 

# Model Definition
R> law <- setLaw_th(method = method_Fourier, up = up, low = low, N = N, 
+    N_grid = N_grid) 
R> regr <- setModel(drift = c("0", "0", "0"), diffusion = matrix("0",3,1), 
+    solve.variable = c("VIX", "EURUSD", "JPYUSD"), xinit = c(0,0,0)) 
Mod <- setLRM(unit_Levy = law, yuima_regressors = regr, LevyRM = "SP")

# Estimation
R> lower <- list(mu1 = -100, mu2 = -200, mu3 = -100,  sigma0 = 0.01)
R> upper <- list(mu1 = 100, mu2 = 100, mu3 = 100, sigma0 = 200.01)
R> start <- list(mu1 = runif(1, -100, 100), mu2 = runif(1, -100, 100), 
+    mu3 = runif(1, -100, 100), sigma0 = runif(1, 0.01, 100))
R> est <- estimation_LRM(start = start, model = Mod, data = yData, upper = upper,  
+    lower = lower, PT = floor(tail(index(Data_eq)/30,1L)/2))
R> summary(est)
\end{CodeInput}
\begin{CodeOutput}
Quasi-Maximum likelihood estimation

Call:
estimation_LRM(start = start, model = Mod, data = yData, upper = upper, 
    lower = lower, PT = floor(tail(index(Data_eq)/30, 1L)/2))

Coefficients:
           Estimate  Std. Error
mu1     0.000335328 0.004901559
mu2    -0.062235188 0.033027232
mu3     0.016291220 0.039382559
sigma0  0.082031691 0.004859138
nu      3.062728822 0.616466913

-2 log L: -1506.067 48.17592 
\end{CodeOutput}
\end{CodeChunk}

\section{Conclusion \label{concl}}

In this paper, we have presented classes and methods for a $t$-L\'evy regression model. In particular, the simulation and the estimation algorithm have been introduced from a computational point of view. Moreover three different algorithms have been developed for computing the density, the cumulative distribution and the quantile functions and the random number generator of the $t$-L\'evy increments defined over a non-unitary interval. These latter methods can be also used in any stochastic model available in \yuima for the definition of the underlying noise. 

\tcb{Based on our simulated data, for the estimation of the degrees of freedom the Laguarre method seems to be more accurate. However, due to the restriction on the number of roots used in the evaluation of the integral, we notice a bias on the estimation of the scale parameter. Such bias seems to be observable also in the other implemented methods when the same number of points is used in the evaluation of the integral for the inversion of the characteristic function. However the COS and the FFT methods, the numerical precision can be improved. In particular, it is possible to ensure that the estimates fall in the asymptotic confidence interval at 95\% level even if the estimates of the degrees of freedom in the simplest model (first example) seem to underestimate the true value. As concerns computational time, the FFT methods with a sufficiently large computational precision ($N\geq 5000$) and $h$ sufficiently small seems to be an acceptable compromise.}

\section*{Acknowledgment}
This work is supported by JST CREST Grant Number JPMJCR2115, Japan and by the MUR-PRIN2022 Grant Number 20225PC98R, Italy CUP codes: H53D23002200006 and G53D23001960006.

\bibliographystyle{plain} 


%
%

%
%

\end{document}